\documentclass[referee]{raa}           

\usepackage{graphicx,times}
\usepackage{natbib}
\usepackage{amssymb,amsmath}
\bibpunct{(}{)}{;}{a}{}{,}
\usepackage{xcolor}

\usepackage[pagebackref=true]{hyperref}

\newcommand{\mum}{\ensuremath{\,\mathrm{\mu m}}}

\begin{document}

\title{The impact of bias row noise to photometric accuracy: case study based on a scientific CMOS detector}
\date{\today}

\volnopage{ {\bf 20XX} Vol.\ {\bf X} No. {\bf XX}, 000--000}
\setcounter{page}{1}

\author{Li Shao\inst{1}, Hu Zhan\inst{1,2}, Chao Liu\inst{1,3,4}, Haonan Chi\inst{1}, Qiuyan Luo\inst{1}, Huaipu Mu\inst{1}, Wenzhong Shi\inst{1}}

\institute{National Astronomical Observatories, Chinese Academy of Sciences, Beijing 100012, China; {\it shaoli@nao.cas.cn}\\
\and
Kavli Institute for Astronomy and Astrophysics, Peking University, Beijing 100871, China \\
\and
University of Chinese Academy of Sciences, Beijing 100049, China \\
\and
Institute for Frontier in Astronomy and Astrophysics, Beijing Normal University, Beijing 100875, China \\
\vs \no
   {\small Received 20XX Month Day; accepted 20XX Month Day}
}

\abstract{We tested a new model of CMOS detector manufactured by the Gpixel Inc, for potential space astronomical application. In laboratory, we obtain some bias images under the typical application environment. In these bias images, clear random row noise pattern is observed. The row noise also contains some characteristic spatial frequencies. We quantitatively estimated the impact of this feature to photometric measurements, by making simulated images. We compared different bias noise types under strict parameter control. The result shows the row noise will significantly deteriorate the photometric accuracy. It effectively increases the readout noise by a factor of 2 to 10. However, if it is properly removed, the image quality and photometric accuracy will be significantly improved.}


\authorrunning{L. Shao et al.}            
\titlerunning{The impact of CMOS bias row noise to photometry}  
\maketitle

%
\section{Introduction}           
\label{sec:intro}

Since the time of digitalization, CCD (charge coupled device) detectors have played an essential role in astronomy \citep{1986ARA&A..24..255M,2001sccd.book.....J,2006hca..book.....H,2015PASP..127.1097L}. But after the invention of CMOS (complementary metal oxide semiconductor) image detector \citep{1997ITED...44.1689F}, CMOS has attracted more and more attentions in astronomy due to its advantages \citep{2013RAA....13..615Q,2014SPIE.9154E..2IW,2016PASP..128c5002M,2017JInst..12C7008J,2020INASR...5..236S,2021arXiv210108144A,2022JATIS...8b6004G,2022JInst..1708004L}, especially in time domain area \citep{2016SPIE.9915E..14P,2022AcASn..63....9N,2022SPIE12180E..4BS}. It also shows the potential to replace CCD \citep[see e.g.][]{2022SPIE12191E..1BG}.

During the development of the Chinese Space Station Telescope (CSST), both CCD and CMOS were considered as possible options for the CSST Survey Camera (CSC). The CSC is designed for large area optical survey. It uses 30 detectors to form a giant 2.3 billion pixel array to cover a field of view of more than one square degree. The optical survey will cover wavelengths ranging from 250 nm to 1000 nm. The detectors are required to have high quantum efficiency, low readout noise, low dark current and large dynamic range to fit for the proposed wide field blind survey \citep{2011SSPMA..41.1441Z,2018cosp...42E3821Z,zhan21}. In order to test whether the state-of-art CMOS detectors can satisfy the scientific requirement, we perform series of tests in the laboratory at the National Astronomical Observatories, Chinese Academy of Sciences, in Beijing. The detector we tested is an HR9090BSI scientific CMOS (sCMOS) detector\footnote{The detector we tested is a customized one. The standard version of this model is officially named as ``GSENSE1081BSI''. A Chinese exoplanet searching project called ``Earth 2.0'' \citep{2022SPIE12180E..4BS} uses the standard version.} manufactured by the Gpixel Inc\footnote{official website: \url{https://www.gpixel.com/}}. This model is under active development in last few years. The Gpixel Inc. also produces some other science level CMOS detectors for (potential) astronomical applications \citep{2020INASR...5..236S,2022SPIE12191E..14G}.

The readout noise is one of the most important parameters to evaluate the performance of a detector. The bias noise properties do have subtle consequences to final photometry or spectroscopy quality \citep{2021PASP..133i4502J}. Only a few studies were done on CMOS detectors in astronomy field \citep{2015JATIS...1c9002B,2016MNRAS.463.4184B,2022JATIS...8b6004G}, probably due to limited application cases. More can be found in non-astronomy scientific applications \citep{2017NatSR...714425D,2021OptL...46..961M}. In this paper, we will study the bias noise properties of this detector and its photometric performance with simulated images. Since this model is still under development, the content of this paper only reflects the status before the end of 2022.

\section{The noise properties of CMOS bias image}
\label{sec:biasprop}

\subsection{The CMOS detector and the bias images}

The bias images were taken from an HR9090BSI sCMOS detector (serial number HR9090BSI-A10LV-SV-22001JC-0027). This one is designed to optimize the performance in visible bands. The pixel physical size is $10\mum\times10\mum$. The detector supports on-chip 16 bit column-parallel analog-to-digital converter (ADC), with up to four low voltage differential signalling (LVDS) channels for image output. The detector uses rolling shutter. The supported minimal exposure time is about $0.6\,\mathrm{ms}$, which is used for taking bias images. The readout time is about $3\,\mathrm{s}$ when using 2 LVDS channels at frequency of 250 MHz. The raw image has $8976\times9222$ pixels. On both left and right sides, there are 34 columns at the edge of array designed as ``electrical black'' (EB) region. The readings of these pixels only come from electronics. In principle, they can be used to calculate bias level as well (similar to overscan region in CCD case). However in practice we found the values in EB region were exceptionally large, and therefore not useful for our analysis. In this paper, we only use the central $8900\times9120$ pixels (``active'' region).

\begin{figure}[ht]
	\begin{center}
		\includegraphics[width=0.9\textwidth]{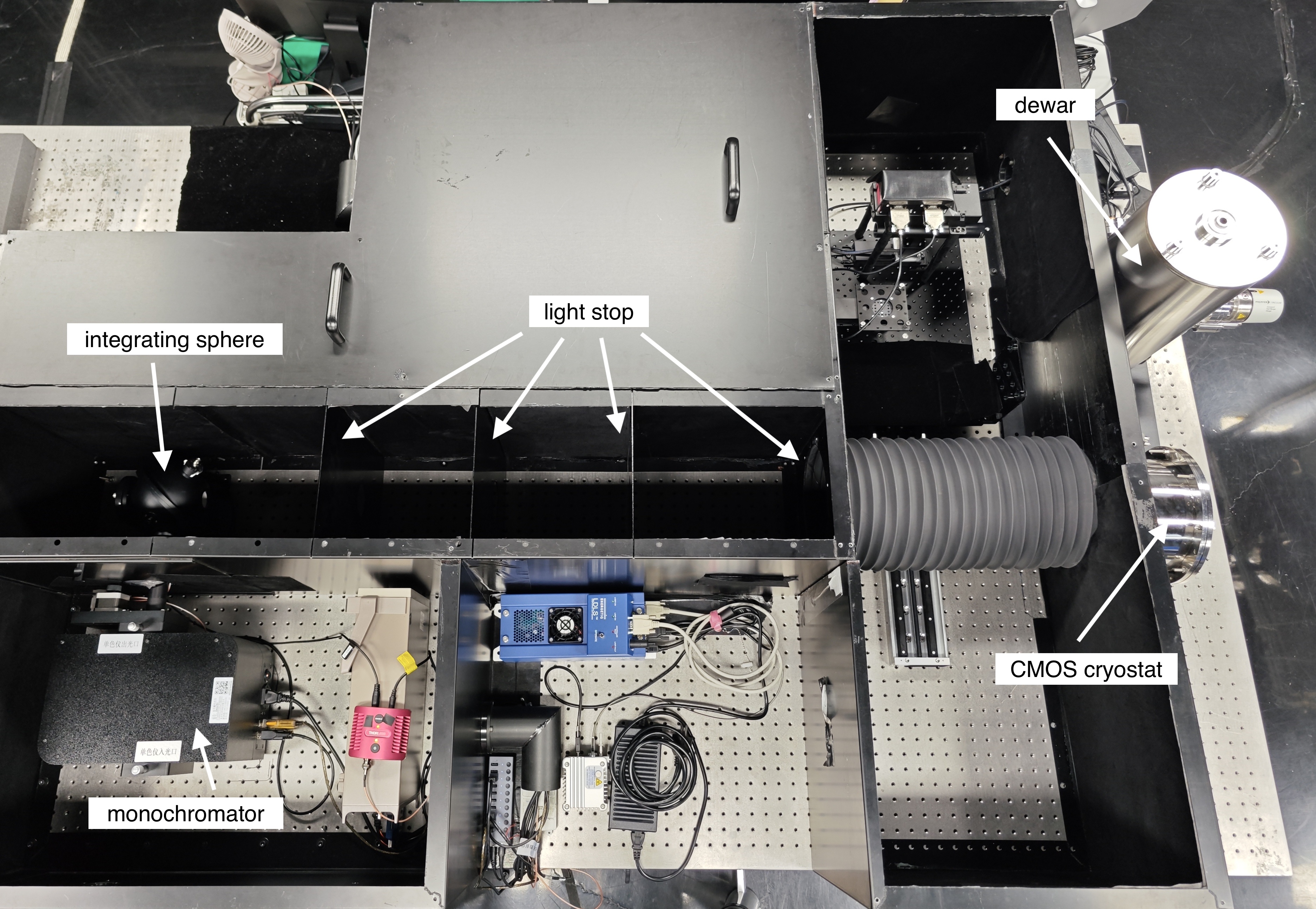}
		\caption{The setup of the detector test bench.}
		\label{fig:testbench}
	\end{center}
\end{figure}

We prepared a test bench (Fig. \ref{fig:testbench}) for the test of CSC detectors in an ISO 7 clean room. It can satisfy multiple test requirements (e.g. pixel response at different wavelengths). The optical system was kept in a light tight box, which was installed on a vibration controlled optical table. All devices emitting light (e.g. running indicator) were carefully separated. Light intensity monitor system was also used when necessary. It makes sure the bias and dark images are not contaminated by any external light source. The detector was installed in a vacuum dewar, cooled with liquid nitrogen. The temperature of the detector was kept at about $188\,\mathrm{K}$, with typical temperature fluctuation no more than $1\,\mathrm{K}$, similar to the designed on-board environment.

The bias pattern is not very sensitive to the temperature. The variation of temperature has negligible contribution to the bias noise. The backend test board was also provided by the Gpixel Inc. (serial number 008). The test was performed in full dark environment to avoid external light contamination to the bias images. Given the typical dark current of $\ngtr0.02\,\mathrm{e^-/s/pixel}$ and high readout frequency, the dark current contribution during readout is negligible. In this paper, we use 30 bias images that were continuously taken on September 5th, 2022. We note that other datasets show similar properties.

In this paper we will use ADU for simplicity. The gain value (in astronomical definition) is about $1.3\,\mathrm{e^-/ADU}$ (measured from photon transfer curve\footnote{In CMOS detector, each pixel has its own gain value. The traditional photon transfer curve technique only gives typical value.}). Readers can convert all the results with this gain if they prefer unit of electrons.

\subsection{Noise residual images and row noise}

We first stacked the 30 bias images to make a master bias. The stacking was performed pixel by pixel. In each pixel, we sorted the 30 values, excluded the top 2 and bottom 2 values to avoid any possible cosmic ray contamination, and then took the average of the rest. We note if we do not exclude any data or exclude more data, the result does not significantly change. On the master bias, we identified a pixel as ``bad'' if its value exceeded the 7-$\sigma$ threshold of typical (median) bias value. There are totally 52565 pixels marked as ``bad'', including 2 rows. Most defected pixels are excluded in this way. The master bias has strong feature of horizontal and vertical lines as well as square blocks. Similar feature was also observed in other CMOS model with large array \citep[e.g.][]{2014SPIE.9154E..2IW,2020SPIE11454E..0GK}.

\begin{figure}[ht]
	\begin{center}
		\includegraphics[width=0.95\textwidth]{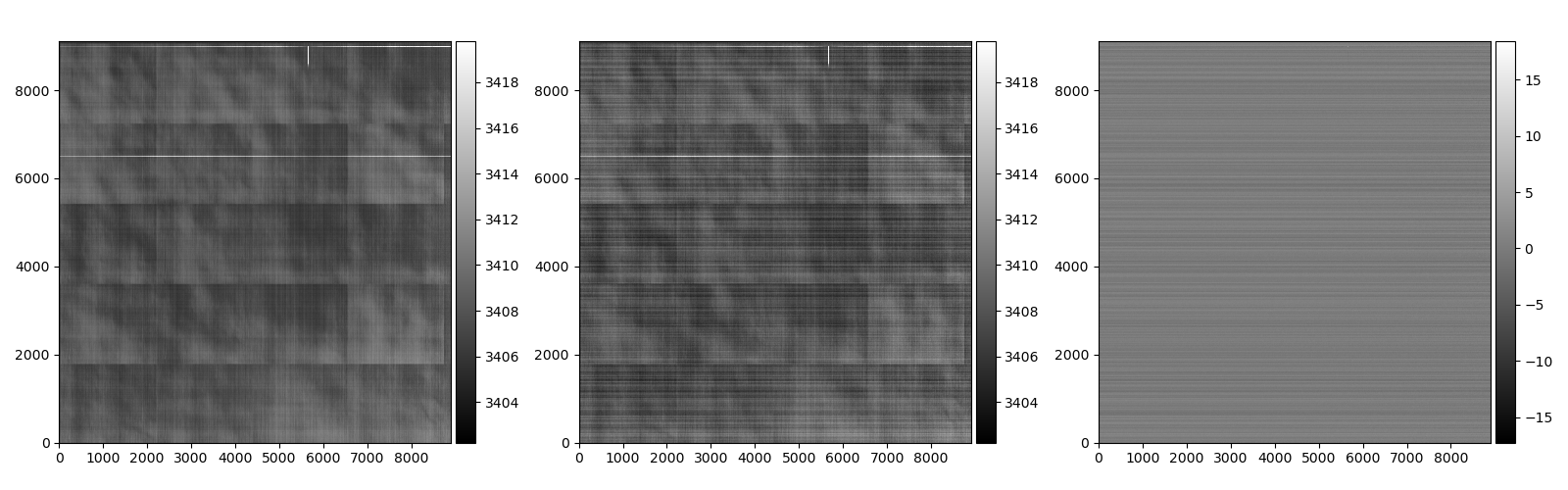}
		\caption{Left: master bias. Middle: individual bias image. Right: noise residual image (middle $-$ left).}
		\label{fig:true_bias}
	\end{center}
\end{figure}

\begin{figure}[ht]
	\begin{center}
		\includegraphics[width=0.9\textwidth]{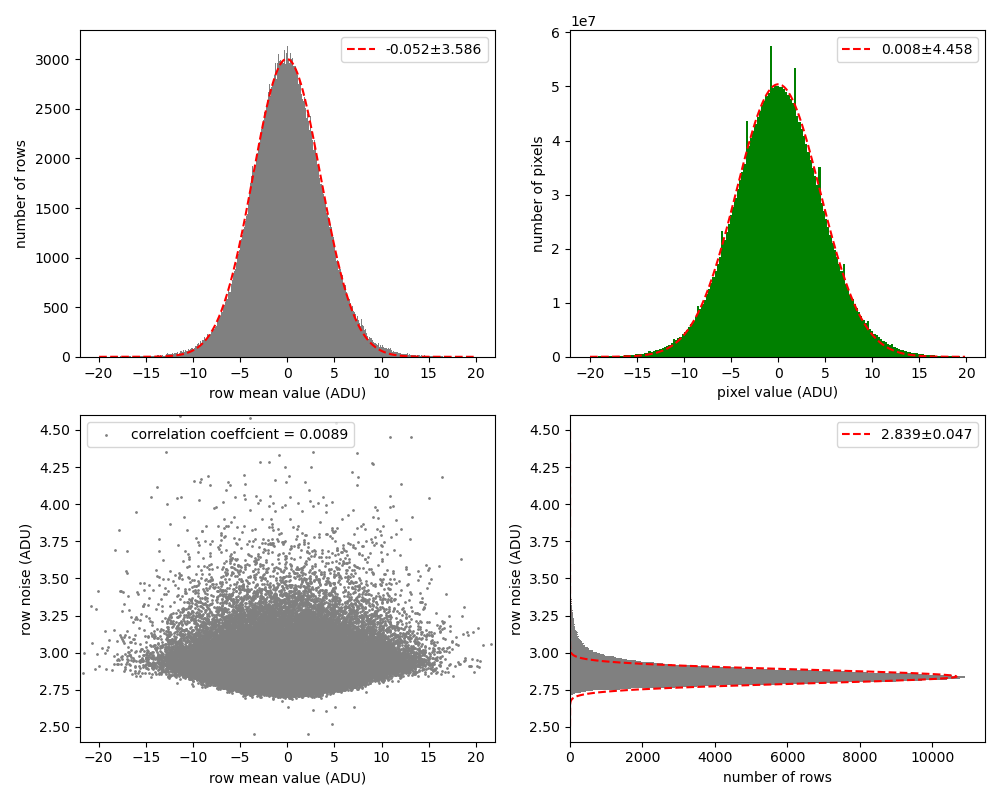}
		\caption{Top right: pixel value histogram of all 30 noise residual images (fixed-pattern-corrected bias images). Bottom left: average value versus standard deviation in each row (the Pearson correlation coefficient is displayed). Top left: histogram of average (in each row) values. Bottom right: histogram of standard deviation (in each row) values. The red dashed line is Gaussian fitting.}
		\label{fig:stats_row}
	\end{center}
\end{figure}

\begin{figure}[ht]
	\begin{center}
		\includegraphics[width=0.9\textwidth]{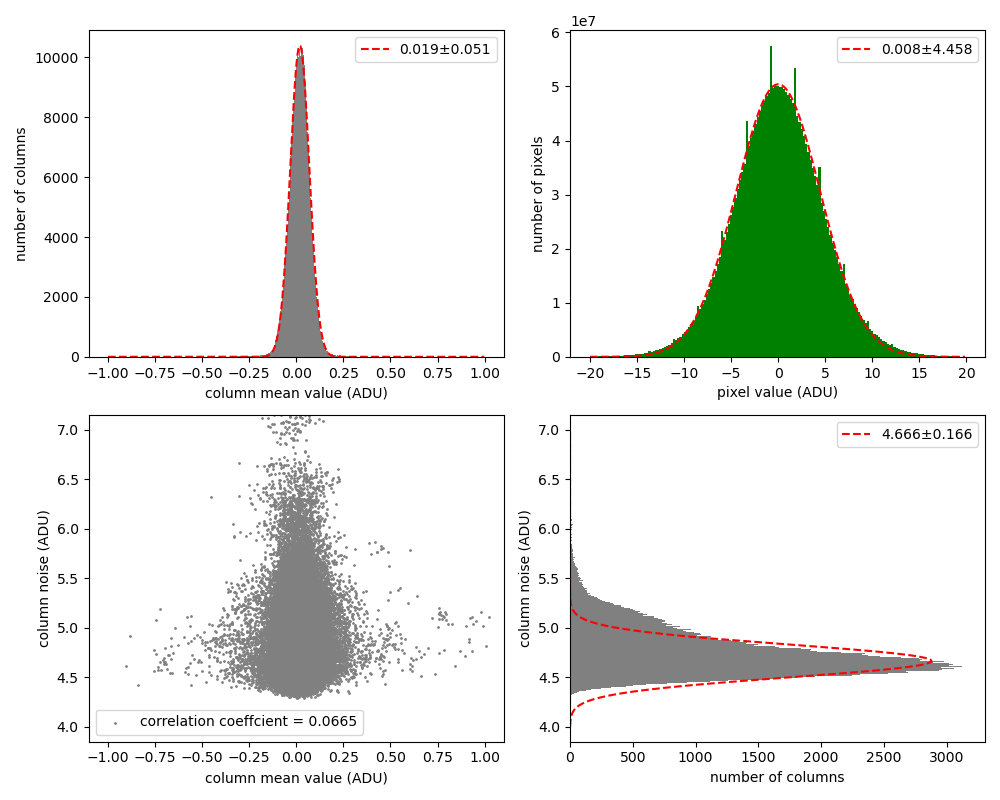}
		\caption{Same as Fig. \ref{fig:stats_row} but in columns.}
		\label{fig:stats_col}
	\end{center}
\end{figure}

The master bias was then subtracted from each individual bias image to remove the fixed pattern, leaving only the noise fluctuation. Hereafter we call such fixed-pattern-corrected bias image as ``{\em noise residual image}''. On the right panel of Fig. \ref{fig:true_bias} there shows one example. The horizontal stripes can be clearly observed. We calculated the average value and the standard deviation in each row or column on the noise residual image. The bad pixels were excluded from the calculation. We merge all the row and column values from the 30 noise residual images together to make statistical plots (Fig. \ref{fig:stats_row} and \ref{fig:stats_col}).

The result shows that the row-to-row variation ($\sigma=3.59$ ADU) is much larger than column-to-column variation ($\sigma=0.05$ ADU). It means the bias images contain spatially correlated noise, which is visually represented as horizontal stripes in noise residual image. The standard deviation in each column is comparable with the pixel value noise (which is based on Gaussian fitting of the pixel histogram, $\sigma=4.46$ ADU) and nominal ``readout noise'' (which is from the standard deviation of noise residual image, $\sigma=4.77$ ADU). Therefore the row-to-row fluctuation is the dominant component of the total readout noise. The standard deviation in each row is actually quite low (typically $\sigma=2.84$ ADU), implying that the detector may have the potential to decrease the overall readout noise to less than 3 ADU.

We notice that the noise distribution is always non-Gaussian. The pixel-wise noise distribution is known to be somehow non-Gaussian. An internal technical report from the Gpixel Inc. (private communication) shows that the histogram of single pixel noise value has a long tail at the high end. Similar behavior was also reported in other CMOS models \citep[e.g.][]{2018Senso..18..977I}.

\begin{figure}[ht]
	\begin{center}
		\includegraphics[width=0.9\textwidth]{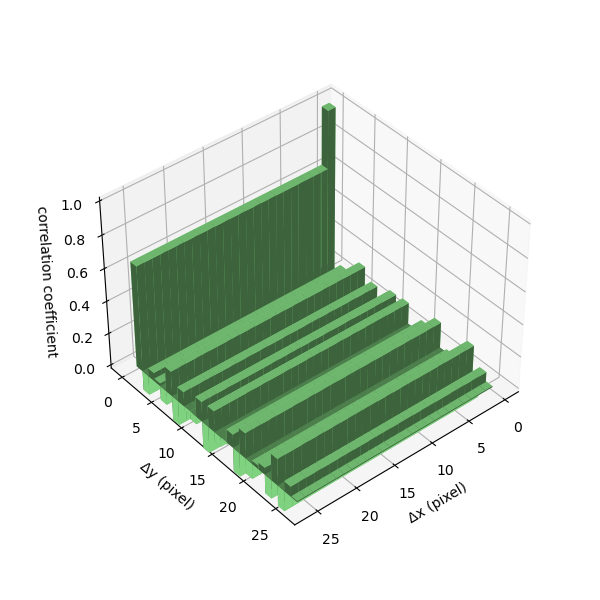}
		\caption{Pixel value correlation coefficient as a function of spatial shift.}
		\label{fig:ccf}
	\end{center}
\end{figure}

We calculated the pixel-to-pixel cross-correlation coefficient as a function of spatial shift (both x and y direction). The result is displayed in Fig. \ref{fig:ccf}. The result clearly shows the correlation in each row (i.e. x direction) is strong ($>0.5$). There are also many horizontal lines with weak but statistically significant correlations, positively or negatively. Hereafter in this paper we call such feature as {\em row noise} ({\em RN}). We note that The pattern also shows some periodicity in y direction, implying the noise also contains characteristic frequency components.

\begin{figure}[ht]
	\begin{center}
		\includegraphics[width=0.95\textwidth]{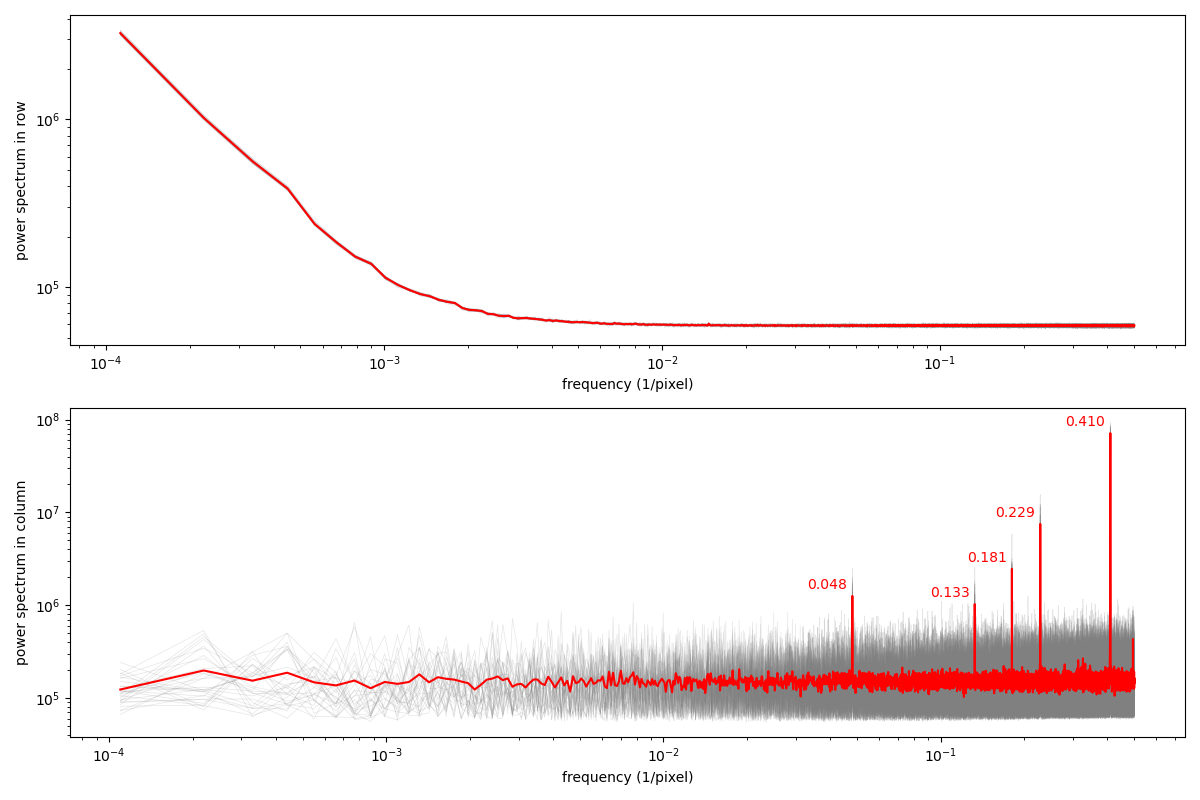}
		\caption{Spatial power spectra of bias noise in rows (top) and columns (bottom). Grey lines are from individual bias images. Red lines are the average of them. In bottom panel, the values of five characteristic frequencies are noted.}
		\label{fig:spec}
	\end{center}
\end{figure}

In order to confirm the existence of the characteristic spatial frequencies in y direction, we calculated the spatial power spectra of the bias noise in Fig. \ref{fig:spec}. The power spectrum is defined as the squared absolute amplitude of the Fourier transform component at given spatial frequency. We calculated the power spectrum in each row (or column) and derived the stacked spectrum with 20\% trimmed mean value for all rows (or columns) in each individual bias image. Then we got 30 spectra (illustrated in gray lines) and derived the average of them (red line). In this way, we can reject any potential outliers (due to bad pixels) to get a robust estimation of the intrinsic power spectrum.

We find that the noise in each row (top panel of Fig. \ref{fig:spec}) has a significant low frequency component, consistent with ``1/f noise'' \citep[see e.g.][]{2000SPIE.3965..168T}. The noise in each column (bottom panel of Fig. \ref{fig:spec}) has flat spectrum except that there are five characteristic frequencies (0.048, 0.133, 0,181, 0.229, 0.410, in unit of $\textrm{pixel}^{-1}$). It may be caused by the power supply fluctuation, which is quite common in sCMOS detectors using column parallel readout architecture \citep[i.e. each row is read in sequence, see][]{Mikkonen2014VerificationOC,2020arXiv201203666G}. The readout could be affected by internal or external electrical circuits which has characteristic frequencies. However, we note that the white noise component is still the dominant noise source (the observed white noise component is equivalent to a noise source with rms of $\nless4$ ADU).

\subsection{Image correction to remove row noise}

The row noise, or more generally stripe noise, is commonly observed in images in remote sensing and/or taken by CMOS detectors \citep[e.g.][]{RSE199515SIMPSON,2007ITGRS..45.1844R,2020arXiv200312751W,2020arXiv201203666G}. There are many developed algorithms to remove the stripe noise, for example, histogram matching with facet filter \citep{2007ITGRS..45.1844R}, midway histogram equalization \citep{2010SPIE.7834E..0ET}, low-rank-based single-image decomposition \citep{IEEE7542167CHANG}, group sparsity based regularization model \citep{NEURO201795CHEN}, wavelet deep neural network \citep{IEEE8678750GUAN}, etc. It is also observed in some astronomical data \citep[e.g.][, for JWST NIRCam in time-series]{2020AJ....160..231S}. The problem is not new and there are many possible solutions. Therefore it will be useful to investigate the properties of row noise corrected images as well.

The origin of row noise in CMOS is complicated \citep{2020arXiv201203666G}. In this paper, only additive (or external signal independent) row noise is discussed (as only bias images are involved)\footnote{Actually, in another imaging test, we find there is row direction crosstalk, i.e. signal dependent row noise, when there are many saturated pixels in a row. But it only appears when there is significantly saturated, i.e. very bright, target. It does not affect the main conclusion of this paper, where we care more about faint objects with low signal-to-noise ratios. If we add the crosstalk in our simulation (presented in next section), the result will never be better.}. Our focus in this paper is not the correction method itself either. Therefore we use a very simple and straightforward method: subtracting the averaged empty pixel value in each row.

\begin{figure}[ht]
	\begin{center}
		\includegraphics[width=0.95\textwidth]{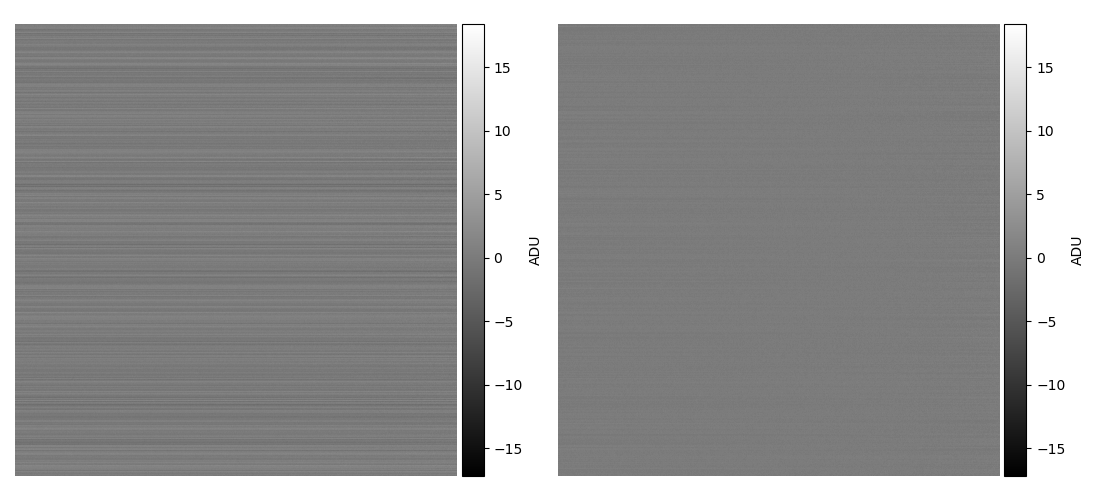}
		\caption{Before (left) and after (right) the row noise correction, displayed in the same scale.}
		\label{fig:correction}
	\end{center}
\end{figure}

To start, it is important to define the empty pixels, i.e. the image area not exposed to external light\footnote{Image area exposed to homogeneously distributed background light but not affected by astronomical objects may also be used for this purpose. However, in real application, such exercise has limitations: e.g. it is difficult for large extended objects (e.g. Magellanic Clouds) or over-crowded region (e.g. star cluster) where hardly any pixel is not contaminated by targets. In this paper, we do not discuss the algorithms to overcome such problems. Nevertheless, if used, it should be tested in real applications. On the other hand, our solution is the simplest way to avoid such problem.}. In case of HR9090BSI, the EB region (34 columns on each side) could be used for this purpose, after proper development in future (suggested by the Gpixel Inc.). Or alternatively, it is possible to mask some active pixels with opaque material to make an optically blind area at the edge of image. This option was proposed by the Gpixel Inc. (private communication) as well. In this paper, we assume the former option will be available. So we test the feasibility of this algorithm here. We use the leftmost 34 and rightmost 34 pixels of the active region (to mimic the size of EB) in each row to calculate the row average. We subtracted the average value of the 68 pixels in each row of the noise residual image. Similar algorithm is often used in CCD image processing, in which the reference empty area is from overscan instead of physical pixels \citep[e.g.][]{2018PASA...35...10W}. In Fig. \ref{fig:correction} the noise residual image before and after the correction are compared. Hereafter, we call these corrected images as ``{\em RN-corr}'' images.

\begin{figure}[ht]
	\begin{center}
		\includegraphics[width=0.9\textwidth]{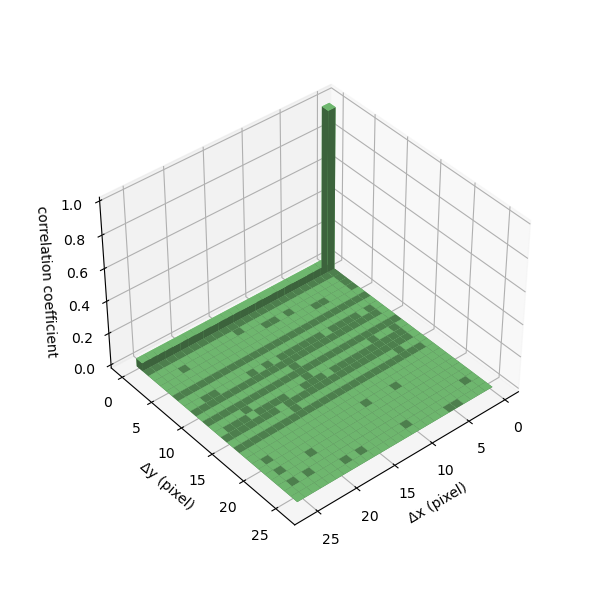}
		\caption{Similar to Fig. \ref{fig:ccf} but for row noise corrected images.}
		\label{fig:ccf_corr}
	\end{center}
\end{figure}

\begin{figure}[ht]
	\begin{center}
		\includegraphics[width=0.95\textwidth]{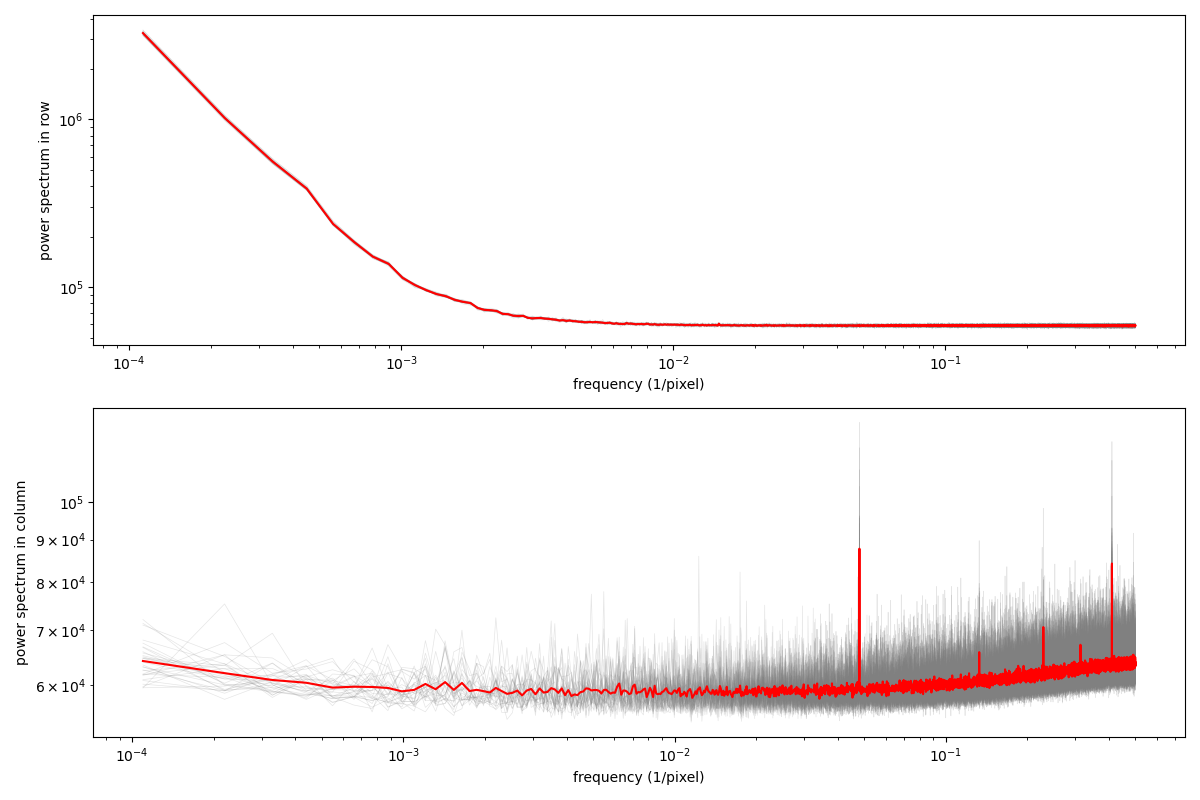}
		\caption{Similar to Fig. \ref{fig:spec} but for row noise corrected images.}
		\label{fig:spec_corr}
	\end{center}
\end{figure}

The row noise correction significantly reduces the readout noise (from 4.77 ADU to 2.89 ADU). The horizontal stripes become much less prominent visually. Similar power spectrum and correlation coefficient analysis were performed on RN-corr images for comparison (see Fig. \ref{fig:ccf_corr} and \ref{fig:spec_corr}).

Although there is still some small pixel-to-pixel correlation residual left in each row, the amplitude is greatly suppressed (Fig. \ref{fig:ccf_corr}). The strength of the power spectrum in vertical direction ($\sim 6\times10^4$, Fig. \ref{fig:spec_corr}) is about one third of the original one ($\sim 2\times10^5$, Fig. \ref{fig:spec}), becoming comparable with the horizontal one. Though as side effect, the power spectrum is no longer flat. Some characteristic frequencies are still there (e.g. at 0.048 $\mathrm{pixel}^{-1}$), but their amplitudes are also greatly reduced. All these results demonstrate that the correction is very helpful. In next section, we will include the corrected images into our analysis as well.

\section{The impact of row noise to photometry}
\label{sec:photometry}

\subsection{The simulated noise residual images}

Here we use simulated images with galaxies and stars to better quantify the effect of row noise in astronomical application. We always simulate master bias corrected images (i.e. object added to the noise residual image).

For RN images, we use cutouts from original CMOS noise residual images. We randomly pick sub-regions of $196\times196$ pixels. Each random block must contain no more than 10 bad pixels and all bad pixels were set to zero. Random blocks should never overlap with each other so that the same noise pattern would not appear twice. There are totally 30446 random cutouts taken from 30 images. We also made cutouts for RN-corr images at the same positions for controlled comparison.

\begin{figure}
	\begin{center}
		\includegraphics[width=0.9\textwidth]{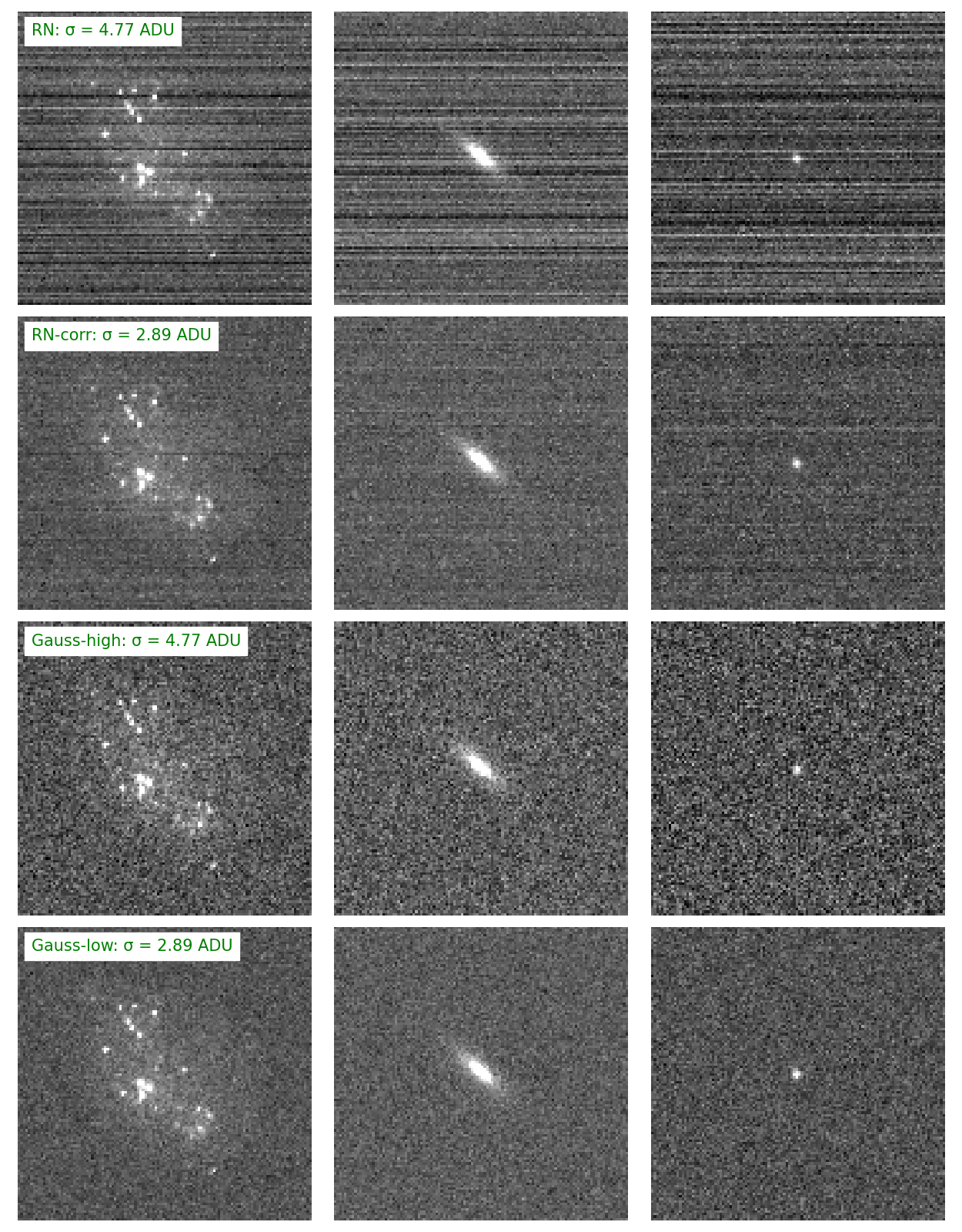}
		\caption{The same object with different noise residual as background (displayed in the same color scale). Left column: real galaxy model; middle column: Sersic model; right column: star model. The images are zoomed in for better visibility.}
		\label{fig:compare_bias}
	\end{center}
\end{figure}

For comparison, we also simulated pure Gaussian noise images. We made two sets of noise residual images with $\sigma=4.77$ ADU ({\em Gauss-high}) and $\sigma=2.89$ ADU ({\em Gauss-low}) respectively. They were used to compare with RN and RN-corr images respectively, in sense of nominal readout noise. We do not only observe the effect of overall readout noise reduction, but also the effect of RN itself. Fig. \ref{fig:compare_bias} compares the different noise residual types.

\subsection{Real galaxy models}

\begin{figure}
	\begin{center}
		\includegraphics[width=0.9\textwidth]{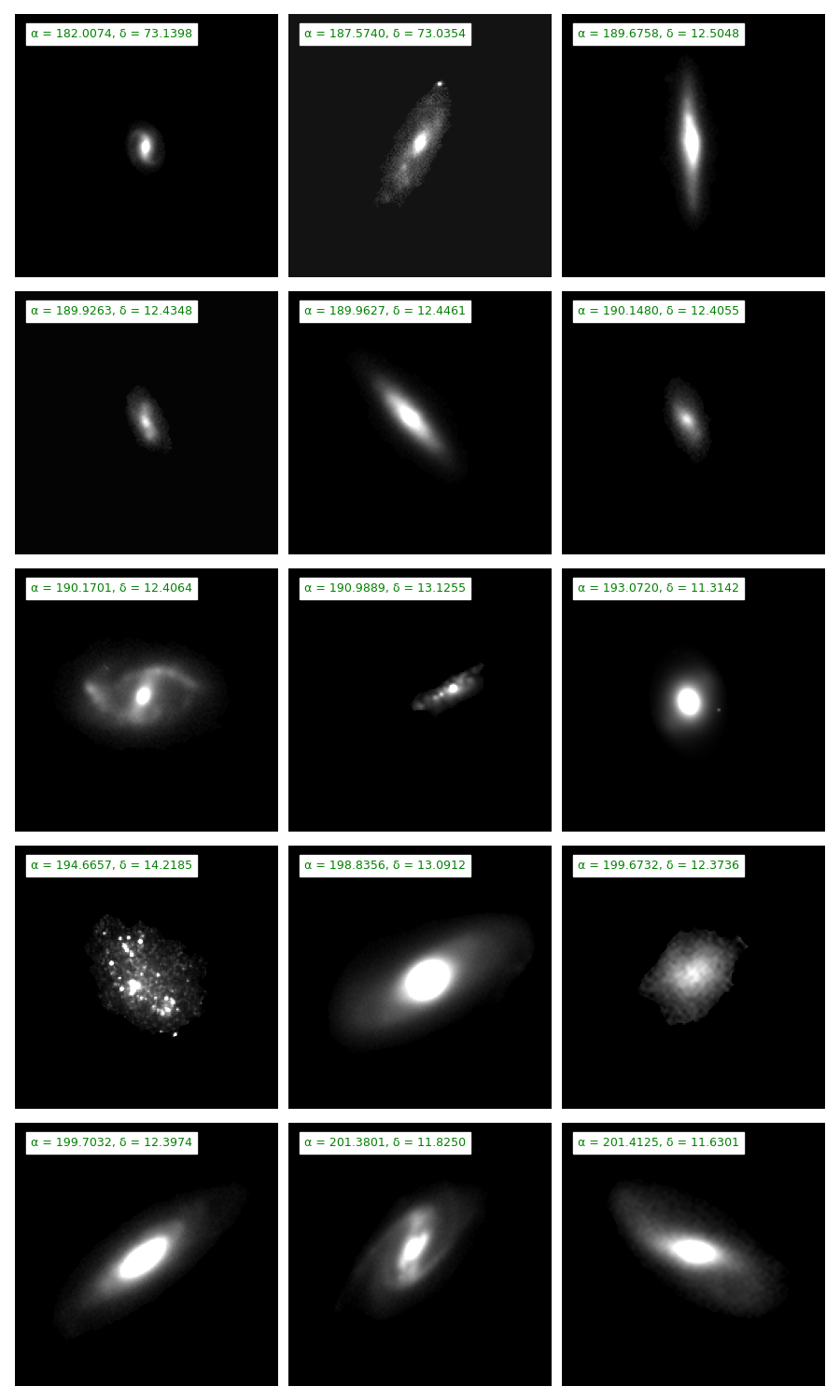}
		\caption{Real galaxy models used in simulation. Right ascension and declination of the image center are shown for each galaxy (some galaxies are not centered).}
		\label{fig:gallary}
	\end{center}
\end{figure}

Real galaxies were added to the simulated noise residual images. The image cutouts of real galaxies were directly taken from the website of the Legacy Survey\footnote{https://www.legacysurvey.org/}\citep{2019AJ....157..168D}. 15 galaxies were randomly picked. For simplicity and better signal-to-noise ratio, all of the three bands were merged together. Source detection and segmentation were performed to remove all nearby neighboring objects (except one galaxy $\alpha=187.5740, \delta=73.0354$ where a possible foreground star were kept to mimic the effect of possible confusion in real pipeline photometry). All pixels out of the target galaxy segment were set to zero. The cutout image were then normalized to unity total flux. Fig. \ref{fig:gallary} shows all the galaxies we used.

When adding the galaxy model to noise residual cutout, we randomly picked one model from the 15 models, multiplied by a random flux scale between 0.1 and 10. We use the same set of models for each type of noise residual image, to make sure the difference only comes from noise residual. The left column of Fig. \ref{fig:compare_bias} show one example of such image set.

\subsection{Sersic models}

\begin{figure}[ht]
	\begin{center}
		\includegraphics[width=0.9\textwidth]{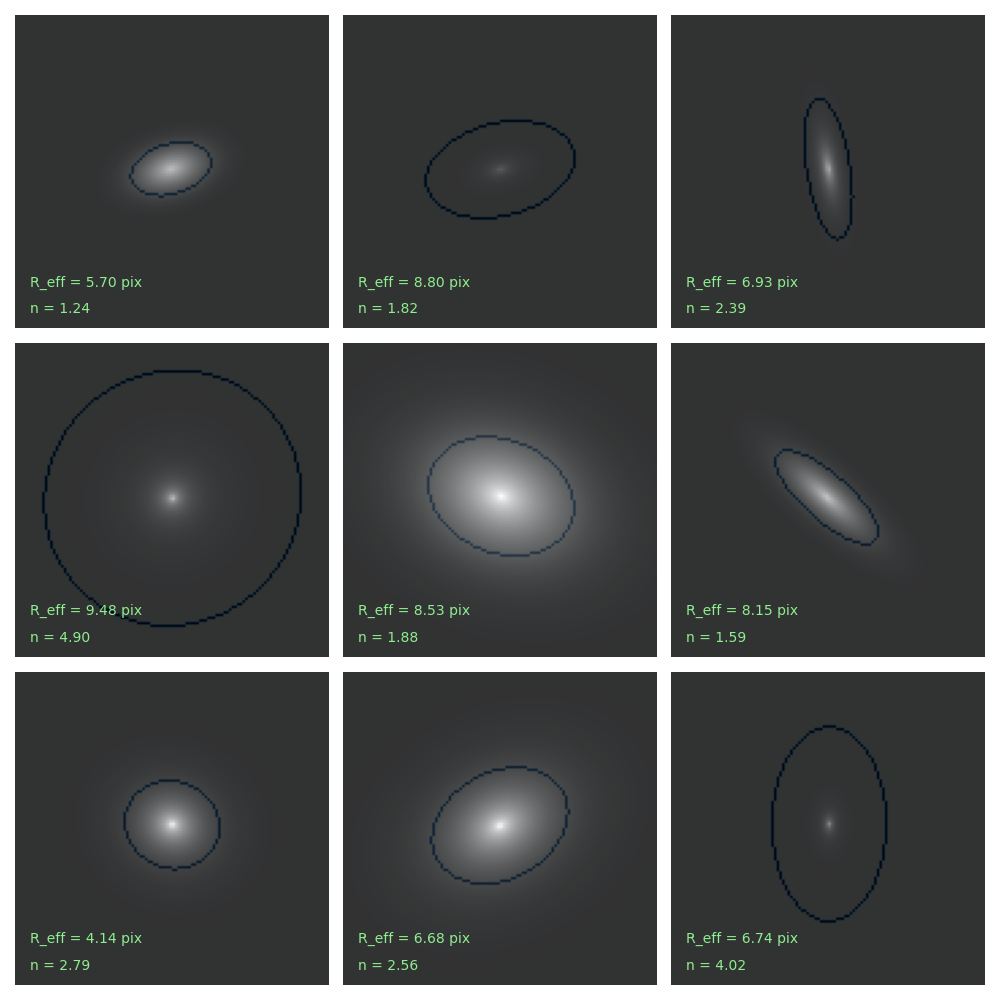}
		\caption{Example Sersic models used in simulation. Blue line shows the segment which contains 90\% of total flux.}
		\label{fig:sersic}
	\end{center}
\end{figure}

Sersic model is a good way to parameterize the light distribution of elliptical galaxies, often used in simplified simulations \citep[e.g.][]{2022MNRAS.515..652H}. We use python package {\it astropy} to make 2D Sersic model images, with random parameters listed in Table \ref{tab:sersic}. Some examples are shown in Fig. \ref{fig:sersic}. Since the Sersic model extends to infinite radius, we made a segment (consistent with an elliptical aperture) which included 90\% of the total flux for further photometric measurement. Different from real galaxy models, Sersic models have larger variability in galaxy size, probably more representative for field galaxies in a blind survey. The middle column of Fig. \ref{fig:compare_bias} show one example for different noise residual background.

\begin{table}
	\begin{center}
		\begin{tabular}{c|c|c}
			Parameter & Min & Max \\ \hline
			amplitude & 100 & $10^5$ \\
			effective radius (pixel) & 3 & 10 \\
			Sersic index & 1 & 5 \\
			ellipticity & 0 & 0.8 \\
			position angle (degree) & 0 & 360 \\
			center x offset (pixel) & -0.5 & 0.5 \\
			center y offset (pixel) & 0.5 & 1.5 \\
		\end{tabular}
	\end{center}
	\caption{Random parameters for Sersic models.}
	\label{tab:sersic}
\end{table}

\subsection{Star models}

To simulate stars, we use 2D Gaussian model with $\sigma=1.13$ pixels (FWHM 2.66 pixels), which is the designed optical resolution (dynamic) of CSC. The model values were first assigned to a 0.05 pixel resolution sub-grid and then summed together to recover the true pixel value. Random amplitudes (between 10 to $10^{3.5}$) and x/y center offsets (between -0.5 and 0.5) were used for each simulated star. Some examples were displayed in Fig. \ref{fig:sersic}. A circular aperture with radius of 4 pixels was used as segment for further photometric measurement. The right column of Fig. \ref{fig:compare_bias} show one example for different noise residual background.

\begin{figure}[ht]
	\begin{center}
		\includegraphics[width=0.9\textwidth]{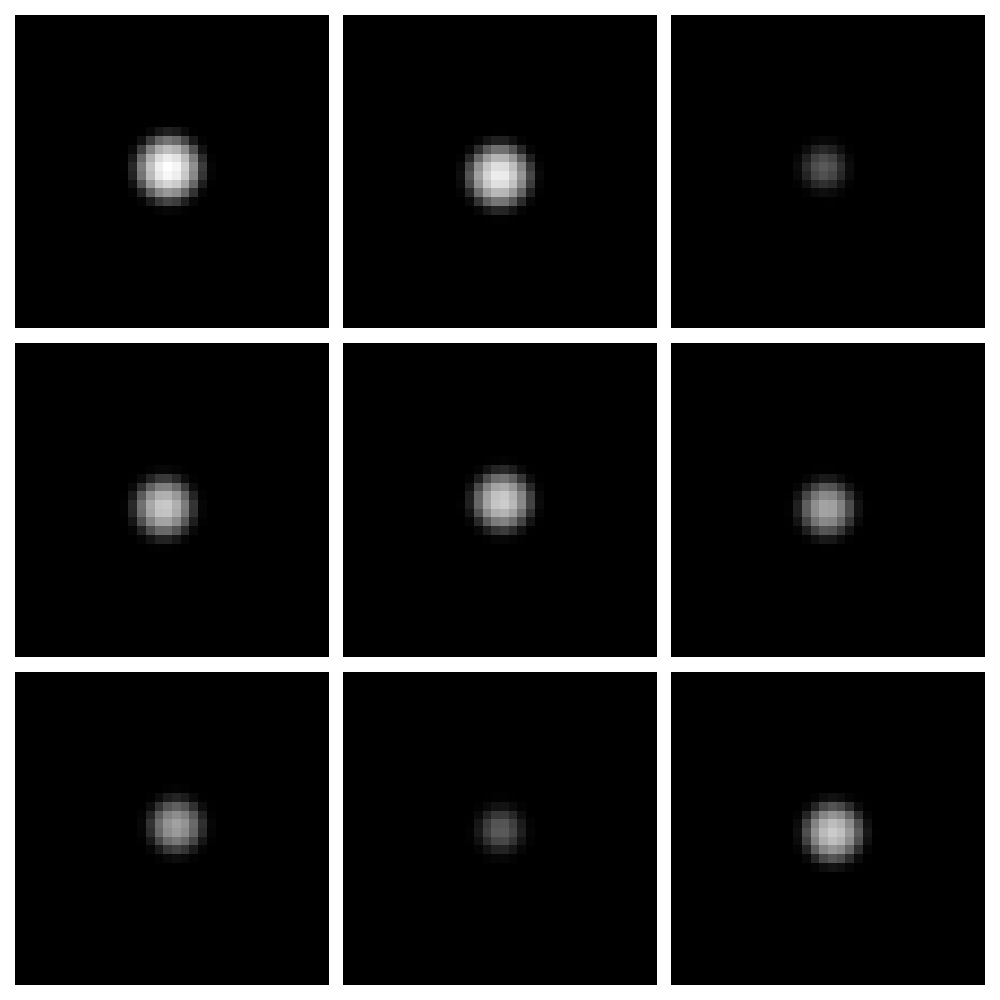}
		\caption{Example Gaussian models used in simulation. All panels are in the same scale. The images are zoomed in.}
		\label{fig:star}
	\end{center}
\end{figure}

\subsection{The photometry and basic parameters}

We used the python package {\it photutils} v1.5 to measure several basic photometric parameters: position, elongation, position angle of the major axis, segment flux, Kron flux and circular aperture flux.

The position measurement is very important in astrometry, i.e. determining the actual position of the object. Many other more complicated photometric parameters are sensitive to position measurement accuracy. The elongation is defined as the ratio between the length of major axis and the length of minor axis. It is often used as an indicator of the inclination angle of spiral galaxy. The position angle of the major axis is defined as the angle between the major axis and y-axis direction. It is also an important parameter to describe how the galaxy is positioned. Both the elongation and position angle are important morphological parameters, especially useful in weak lensing shear measurement (although the actual algorithm is more complicated).

The segment flux is the sum of pixel values within the segment. It is a model-independent flux measurement, which is believed to be a robust estimation of the total flux. The Kron flux \citep{1980ApJS...43..305K} is another parameter to calculate the total flux. It sums up the flux within an elliptical aperture (i.e. assuming the galaxy shape can be represented by an elliptical). The parameters of the ellipse are determined from the first-order and second-order moment values. We use the default parameter in {\it photutils}. Comparing with segment flux, Kron flux is more sensitive to position and morphological measurements, but more reasonable in sense of measuring the total flux of a (generally symmetric) galaxy. It is widely used in many optical surveys \citep[e.g.][]{2013ApJS..207...24G, 2016arXiv161205560C}. The circular aperture flux was also calculated, which was preferred for stars. We always used the circular aperture with radius of 1.5 FWHM (3.99 pixels), which was sufficiently large to include most flux of a point source (more than 99.9\% in our case) and was not too large to include too many noisy empty pixels. Most astrophysical studies require accurate flux measurement. It is important to check the reliability of the flux measurement.

We used the fixed segment, i.e. the segment of the object (either galaxy or star) was fixed to the model image segment mentioned in previous subsections. If we use free segment or use resampled image, the conclusion does not change. We also note that Poisson resampling of the input flux was not used in our simulated images. In this way we make sure the difference {\em only} comes from the bias noise fluctuation within the same area. We will make a direct comparison between different noise types at the same nominal ``readout noise'' level and ``signal-to-noise ratio''.

\subsection{Results}
\label{subsec:hcn_result}

We define the measurement error as the measurement difference between simulated image (model + noise) and pure model image, i.e. we treat the pure model result as ``true value''. If the measurement is flux or elongation, the difference will be normalized by the ``true value'' and presented in percentage.

\begin{figure}
	\begin{center}
		\includegraphics[width=0.9\textwidth]{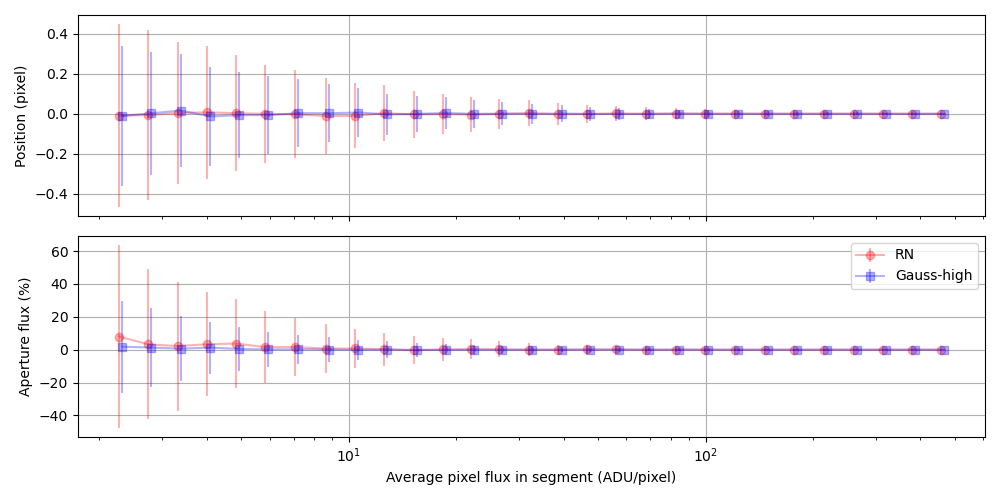}
		\caption{Measurement error as a function of average pixel flux in detection segment (see text) for simulated stars. The blue data points are slightly shifted along X axis for better visibility.}
		\label{fig:br_value_star}
	\end{center}
\end{figure}

Fig. \ref{fig:br_value_star} shows the measurement error as a function of average pixel flux in segment, which is defined as the total flux (in ADU) within the detection segment divided by the number of pixels of the segment. Here the position offset is defined as the absolute position offset with the sign of Y offset (i.e. $|\Delta Y| / {\Delta Y} \times \sqrt{\Delta X^2 + \Delta Y^2}$). Each data point is binned from 1000 measurements. Error bars show the variation within the bin. Statistically, the mean value is accurate even for RN. But at any input brightness, the uncertainty of RN measurements are clearly worse than the Gaussian counterparts (note they have the same ``readout noise''). Real galaxies and Sersic models have similar results.

\begin{figure}
	\begin{center}
		\includegraphics[width=0.9\textwidth]{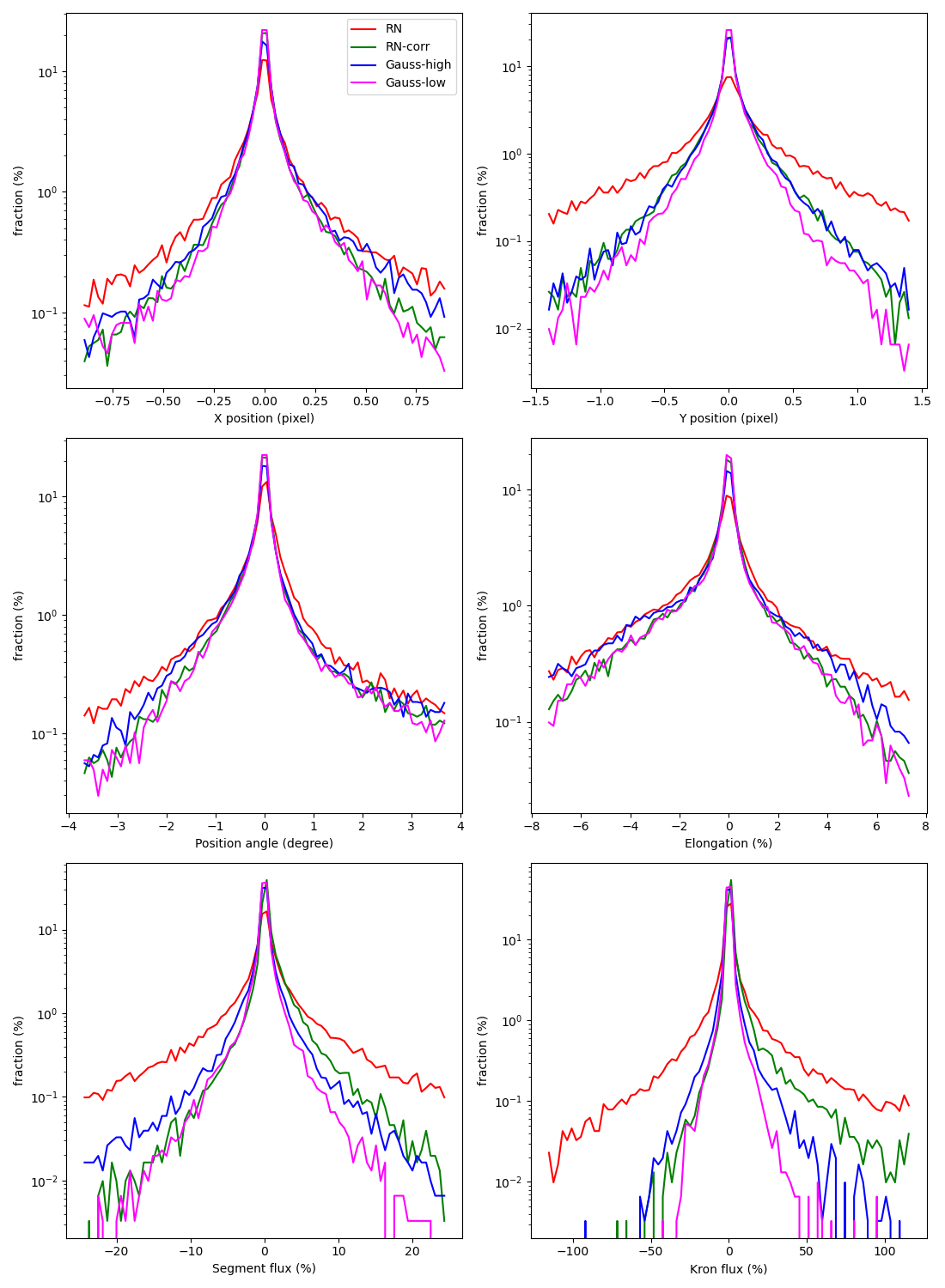}
		\caption{The distribution of measurement errors for real galaxy models. Different bias noise types are displayed in different colors.}
		\label{fig:hist_galaxy}
	\end{center}
\end{figure}

Fig. \ref{fig:hist_galaxy} shows the distribution of measurement error of all real galaxy images for 6 parameters. It shows the performance of RN is the worst for all parameters. The measurement error is especially large for Y position, segment flux and Kron flux. After correction, the performance of RN-corr is generally between Gauss-high and Gauss-low cases. It suggests the row noise correction is not perfect but still very useful.

The distribution is sometimes not symmetric. It may be related to the underlying algorithm and properties of the input galaxy templates. Such asymmetry does not affect our conclusion. Interestingly, the Y position uncertainty in RN case is significantly worse than X position. It is somehow expected, given the anisotropic nature of row noise.

\begin{figure}
	\begin{center}
		\includegraphics[width=0.9\textwidth]{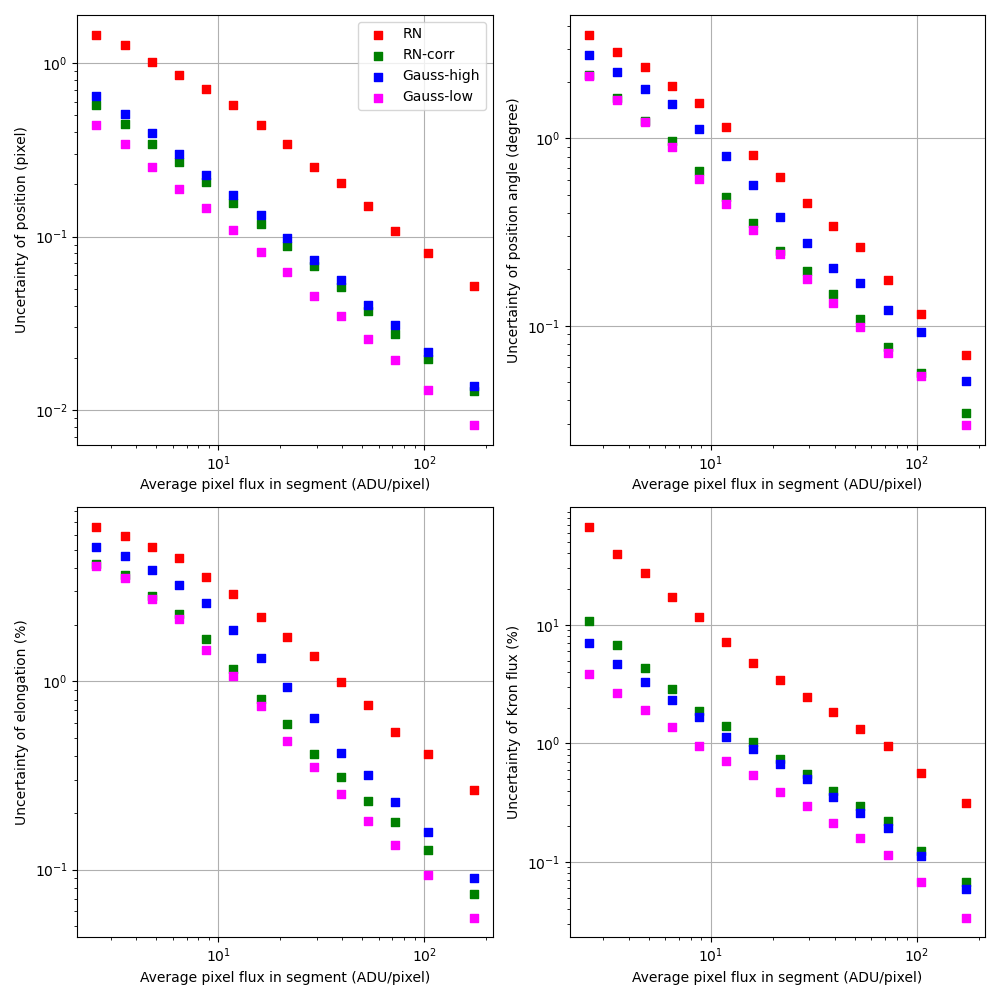}
		\caption{The measurement uncertainty as a function of average pixel flux in detection segment for real galaxy models. Different bias noise types are displayed in different colors.}
		\label{fig:br_uncert_galaxy}
	\end{center}
\end{figure}

\begin{figure}
	\begin{center}
		\includegraphics[width=0.9\textwidth]{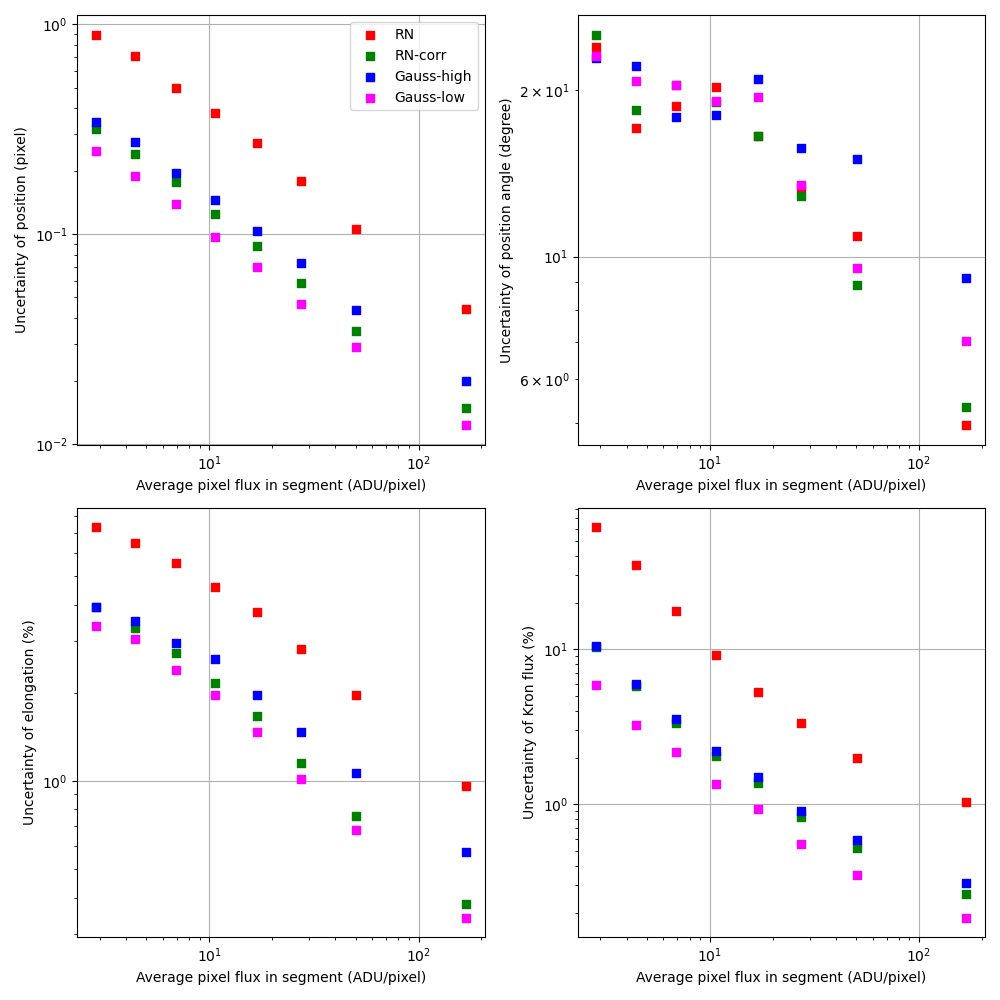}
		\caption{Same as Fig. \ref{fig:br_uncert_galaxy} but for simulated Sersic images.}
		\label{fig:br_uncert_sersic}
	\end{center}
\end{figure}

\begin{figure}
	\begin{center}
		\includegraphics[width=0.9\textwidth]{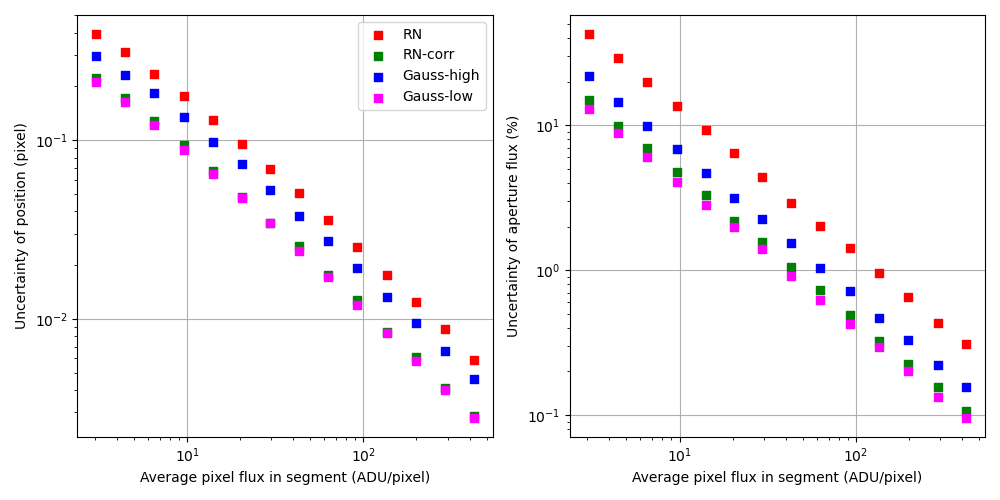}
		\caption{Same as Fig. \ref{fig:br_uncert_galaxy} but for simulated stars.}
		\label{fig:br_uncert_star}
	\end{center}
\end{figure}

To better quantify the difference between different noise types, we plot the measurement uncertainty (the standard deviation of measurement error within the bin) as a function of average pixel flux in segment. Fig. \ref{fig:br_uncert_galaxy} to \ref{fig:br_uncert_star} show the results of all 4 noise types combined with 3 different models.

The result is similar. Except the position angle of Sersic images, RN is significantly worse than all other cases. RN-corr is much improved, comparable with Gaussian (between the high and low noise cases). The effect is much prominent in galaxies (including real galaxies and Sersic models) than stars. It is probably due to the fact that the flux measurement apertures of stars are usually much smaller than that of galaxies. More pixels means higher noise contribution from bias noise, and hence more vulnerable to row noise. It also clearly demonstrate that the correction is useful but the corrected result is still not as good as pure white noise.

By comparing the flux uncertainties with the same input brightness, it is found that the uncertainty of RN case is about 3 to 10 times larger than Gauss-high for galaxies, and about 2 times larger for stars. In this paper, our test strictly controls the variables, leaving only the bias noise as the only parameter. So the effect of row noise we measured here can be directly translated into the increase of {\em effective readout noise}, by a factor of 2 to 10, depending on target type. If we compare RN with Gauss-low, the difference is larger.

\section{Discussion and summary}
\label{sec:conclusion}

In this paper, we studied the noise properties of bias images taken from an HR9090BSI sCMOS detector. We found a strong pixel cross-correlation in each row, i.e. the row noise, and some minor fluctuations with fixed vertical spatial frequencies in column direction. A simple row noise correction was applied to the real bias image, leading to a much lower readout noise and much smaller pixel-to-pixel correlation. We made a set of simulated galaxies and stars with different noise residual images as background. A series of photometry tests were performed. The results suggest that the row noise feature can significantly deteriorate the overall photometry performance, especially for galaxies. It {\em effectively} increases the readout noise by a factor of 2 to 10, depending on the type of object.

We demonstrate that the existence of row noise pattern, i.e. pixel-to-pixel correlation, could greatly increase the photometric uncertainty. Such effect is not simply due to the increase of nominal readout noise. In fact, at the same nominal readout noise level, the row noise case performs much worse than its Gaussian noise counterpart. It again proves the importance of pixel-to-pixel independence in astronomical high-precision photometry. It also makes a clear warning to anybody working on observation planning or camera design. The nominal ``readout noise'' may be misleading when the detector shows strong pixel-to-pixel correlation (no matter what the origin of such correlation is). It may underestimate the required observation time for a given photometric accuracy, leading to an ambiguous observation result, or overestimate the performance of camera, leading to a failure of achieving the original design goal.

We also note that, given the fact the row noise is very common in sCMOS detectors, it is necessary to seriously consider the destriping method, either implemented in hardware or adopted at the data processing stage with efficient and robust algorithm. Our result suggests even the simplest destriping algorithm can make prominent improvement. Any future astronomical project using sCMOS detector should be aware of these facts.

Our analysis also has limitations, although we note that some of them are out of the scope of this paper. First, our analysis of row noise is totally phenomenal. We treat the camera as a black box (due to some reasons). In order to fully understand the origin of row noise in this specific CMOS model, we will need more knowledge of the detector system, including the manufacturing details of the chip and the readout electronics. Second, we only consider the additive signal-independent row noise. Other multiplicative and/or signal-dependent noise may also affect the photometry. Of course, all these additional noise components will further worsen the photometry result if not properly corrected. In order identify these noise components, we will need more data, especially images from real astronomical observations. We note that in near future the ``Earth 2.0'' project \citep{2022SPIE12180E..4BS} may give us a good chance. Third, this paper is focused on bias noise only. Our result is a good quantitative evaluator of the image readout performance. However, as a side effect, it makes the result looks too ``exaggerated'' when comparing to real situation. In practice, the bias noise usually only contributes a small fraction of the total photometric uncertainty. Many noise sources, such as Poisson noise, dark current, sky background and stray light, will make major contributions. Readers should be cautious when interpreting our results. To make a more realistic analysis of the photometric uncertainties, data from a more sophisticated simulation (e.g. Fang et al., in preparation, specifically designed for CSST\footnote{CSST simulation v2.0 release: \url{https://csst-tb.bao.ac.cn/code/csst_sim/csst-simulation/-/tree/release_v2.0}}) should be used. Finally, we do not discuss too much about the algorithm for row noise correction. When more and more CMOS data appears in the field of optical astronomy, it may be interesting to systematically study the effect of existing methods to high-precision photometry based on real observation images.

\normalem
\begin{acknowledgements}

We thank the anonymous referee for her/his comments which help to improve this paper.

This work is support by the National Key R\&D Program of China No. 2022YFF0503400.

The Legacy Surveys consist of three individual and complementary projects: the Dark Energy Camera Legacy Survey (DECaLS; Proposal ID \#2014B-0404; PIs: David Schlegel and Arjun Dey), the Beijing-Arizona Sky Survey (BASS; NOAO Prop. ID \#2015A-0801; PIs: Zhou Xu and Xiaohui Fan), and the Mayall z-band Legacy Survey (MzLS; Prop. ID \#2016A-0453; PI: Arjun Dey). DECaLS, BASS and MzLS together include data obtained, respectively, at the Blanco telescope, Cerro Tololo Inter-American Observatory, NSF’s NOIRLab; the Bok telescope, Steward Observatory, University of Arizona; and the Mayall telescope, Kitt Peak National Observatory, NOIRLab. Pipeline processing and analyses of the data were supported by NOIRLab and the Lawrence Berkeley National Laboratory (LBNL). The Legacy Surveys project is honored to be permitted to conduct astronomical research on Iolkam Du’ag (Kitt Peak), a mountain with particular significance to the Tohono O’odham Nation.

NOIRLab is operated by the Association of Universities for Research in Astronomy (AURA) under a cooperative agreement with the National Science Foundation. LBNL is managed by the Regents of the University of California under contract to the U.S. Department of Energy.

This project used data obtained with the Dark Energy Camera (DECam), which was constructed by the Dark Energy Survey (DES) collaboration. Funding for the DES Projects has been provided by the U.S. Department of Energy, the U.S. National Science Foundation, the Ministry of Science and Education of Spain, the Science and Technology Facilities Council of the United Kingdom, the Higher Education Funding Council for England, the National Center for Supercomputing Applications at the University of Illinois at Urbana-Champaign, the Kavli Institute of Cosmological Physics at the University of Chicago, Center for Cosmology and Astro-Particle Physics at the Ohio State University, the Mitchell Institute for Fundamental Physics and Astronomy at Texas A\&M University, Financiadora de Estudos e Projetos, Fundacao Carlos Chagas Filho de Amparo, Financiadora de Estudos e Projetos, Fundacao Carlos Chagas Filho de Amparo a Pesquisa do Estado do Rio de Janeiro, Conselho Nacional de Desenvolvimento Cientifico e Tecnologico and the Ministerio da Ciencia, Tecnologia e Inovacao, the Deutsche Forschungsgemeinschaft and the Collaborating Institutions in the Dark Energy Survey. The Collaborating Institutions are Argonne National Laboratory, the University of California at Santa Cruz, the University of Cambridge, Centro de Investigaciones Energeticas, Medioambientales y Tecnologicas-Madrid, the University of Chicago, University College London, the DES-Brazil Consortium, the University of Edinburgh, the Eidgenossische Technische Hochschule (ETH) Zurich, Fermi National Accelerator Laboratory, the University of Illinois at Urbana-Champaign, the Institut de Ciencies de l’Espai (IEEC/CSIC), the Institut de Fisica d’Altes Energies, Lawrence Berkeley National Laboratory, the Ludwig Maximilians Universitat Munchen and the associated Excellence Cluster Universe, the University of Michigan, NSF’s NOIRLab, the University of Nottingham, the Ohio State University, the University of Pennsylvania, the University of Portsmouth, SLAC National Accelerator Laboratory, Stanford University, the University of Sussex, and Texas A\&M University.

BASS is a key project of the Telescope Access Program (TAP), which has been funded by the National Astronomical Observatories of China, the Chinese Academy of Sciences (the Strategic Priority Research Program “The Emergence of Cosmological Structures” Grant \# XDB09000000), and the Special Fund for Astronomy from the Ministry of Finance. The BASS is also supported by the External Cooperation Program of Chinese Academy of Sciences (Grant \# 114A11KYSB20160057), and Chinese National Natural Science Foundation (Grant \# 12120101003, \# 11433005).

The Legacy Survey team makes use of data products from the Near-Earth Object Wide-field Infrared Survey Explorer (NEOWISE), which is a project of the Jet Propulsion Laboratory/California Institute of Technology. NEOWISE is funded by the National Aeronautics and Space Administration.

The Legacy Surveys imaging of the DESI footprint is supported by the Director, Office of Science, Office of High Energy Physics of the U.S. Department of Energy under Contract No. DE-AC02-05CH1123, by the National Energy Research Scientific Computing Center, a DOE Office of Science User Facility under the same contract; and by the U.S. National Science Foundation, Division of Astronomical Sciences under Contract No. AST-0950945 to NOAO.

\end{acknowledgements}

\bibliographystyle{raa}
\bibliography{bibtex}

\end{document}